\documentclass[preprint,showpacs,preprintnumbers,amsmath,amssymb]{revtex4}
\usepackage{graphicx,epsfig}
\usepackage{amssymb}
\usepackage{amsmath}

\usepackage{graphicx}
\usepackage{dcolumn}
\usepackage{bm}
\usepackage{epsfig}
\usepackage{graphics}

\begin{document}


\title{Bound states of scalar particles in the presence of a short
range potential}

\author{Luis A. Gonz\'alez-D\'{\i}az.}
\affiliation{Centro de F\'{\i}sica IVIC Apdo 21827, Caracas 1020A,
Venezuela
}%
\author{V\'{\i}ctor M. Villalba}
\email{villalba@ivic.ve} \affiliation{Centro de F\'{\i}sica IVIC
Apdo 21827, Caracas 1020A, Venezuela
}%

\date{\today}

\begin{abstract}
We analyze the behavior of the energy spectrum of the Klein-Gordon
equation in the presence of a truncated hyperbolic tangent
potential. From our analysis we obtain that, for some values of the
potential there is embedding of the bound states into the negative
energy continuum, showing that, in opposition to the general belief,
relativistic scalar particles in one-dimensional short range
potentials can exhibit resonant behavior and not only the
Schiff-Snyder effect.
\end{abstract}


\maketitle

The study of relativistic scalar particles in the presence of strong
electromagnetic and gravitational fields is  a topic that has been
carefully discussed in the literature\cite{Greiner1,Greiner}, mainly
because their implications in quantum  phenomena like superradiance
and black hole evaporation\cite{Fulling} as well as their
astrophysical implications in the study of boson stars. Recently
research has been conducted on  quantum effects of Klein-Gordon
particles in the presence of strong magnetic
fields\cite{Khalilov,Khalilov2}. The pioneering works on
Klein-Gordon particles (pion, etc) date back to 1940 when Schiff,
Snyder and Weinberg solved the Klein-Gordon equation for a
square-well potential. A surprising outcome of this paper was that
for strong potentials antiparticles states emerge from the lower
continuum. A systematic study of the solutions of the Klein-Gordon
equation for various types of potentials, aiming to investigate the
conditions necessary for antiparticle binding,  were performed by
Fleischer and Soff\cite{Fleischer}.

Quantum effects associated with scalar particles in strong
short-range electric fields exhibit some particularities that
establish a fundamental difference between the eigenvalues and phase
shifts for Dirac particles and Klein-Gordon particles when the
intensity of the external potential surpasses the supercritical
value. The bound $s$ state associated with the Klein-Gordon equation
does not disappear and it does not behave like a resonance in the
lower continuum of the energy spectrum. Such a state becomes a
complex eigenstate with zero norm. Bawin and Lavine\cite{BL:79} have
shown that, for $p$ waves the eigenvalue becomes complex without the
appearance of the Schiff - Snyder - Weinberg effect, phenomenon that
takes place for $s$ waves and consists in the coalescence of
particle and antiparticle eigenstates before they become
complex\cite{SSW:40,Weinberg}. It is worth mentioning that, since
the Schiff-Snyder effect is a particular effect associated with
short-range interacting potentials,  it is not observed in the
presence of Coulomb interactions\cite{Rafelski}

In this article, using
the phase shift approach \cite{Newton} and the Wigner time
delay \cite{EW:55}, we show that the phenomenon described in Ref.
\cite{BL:79} also takes place for $s$ waves in the presence of a
radial hyperbolic tangent potential. The potential is different 
from zero in the range $[0,a]$, with $a>0$ transforming the potential 
into a
short-range potential. For this potential,  with a judicious choice
of the parameters (Fig. 1), we observe the phenomenon of resonance, which
manifests itself via the the embedding in the lower negative
continuum.  This situation becomes a counterexample to the general
result shown by Popov \cite{VP:71} according to which, for
short-range potentials, $s$ waves solutions always exhibit the
Schiff-Snyder-Weinberg effect and therefore there is no embedding
of the bound states into the negative continuum.

The radial equation for $s$ waves is given by \cite{Greiner}
\begin{equation}\label{kg}
\frac{d^{2}u(r)}{dr^{2}}+\left(\left(E-V(r)\right)^2-m^2\right)u(r)=0
\end{equation}
where $eA_{0}=V(r)$ with
\begin{equation}
V(r)=\begin{cases} D\tanh(kr)-B,&\,r\in[0,a],\\
0,&\,r\in[a,\infty). \label{V}
\end{cases}
\end{equation}
and we have adopted the natural units $\hslash=c=1$,
\begin{figure}[tbp]
\begin{center}
\includegraphics[width=10cm]{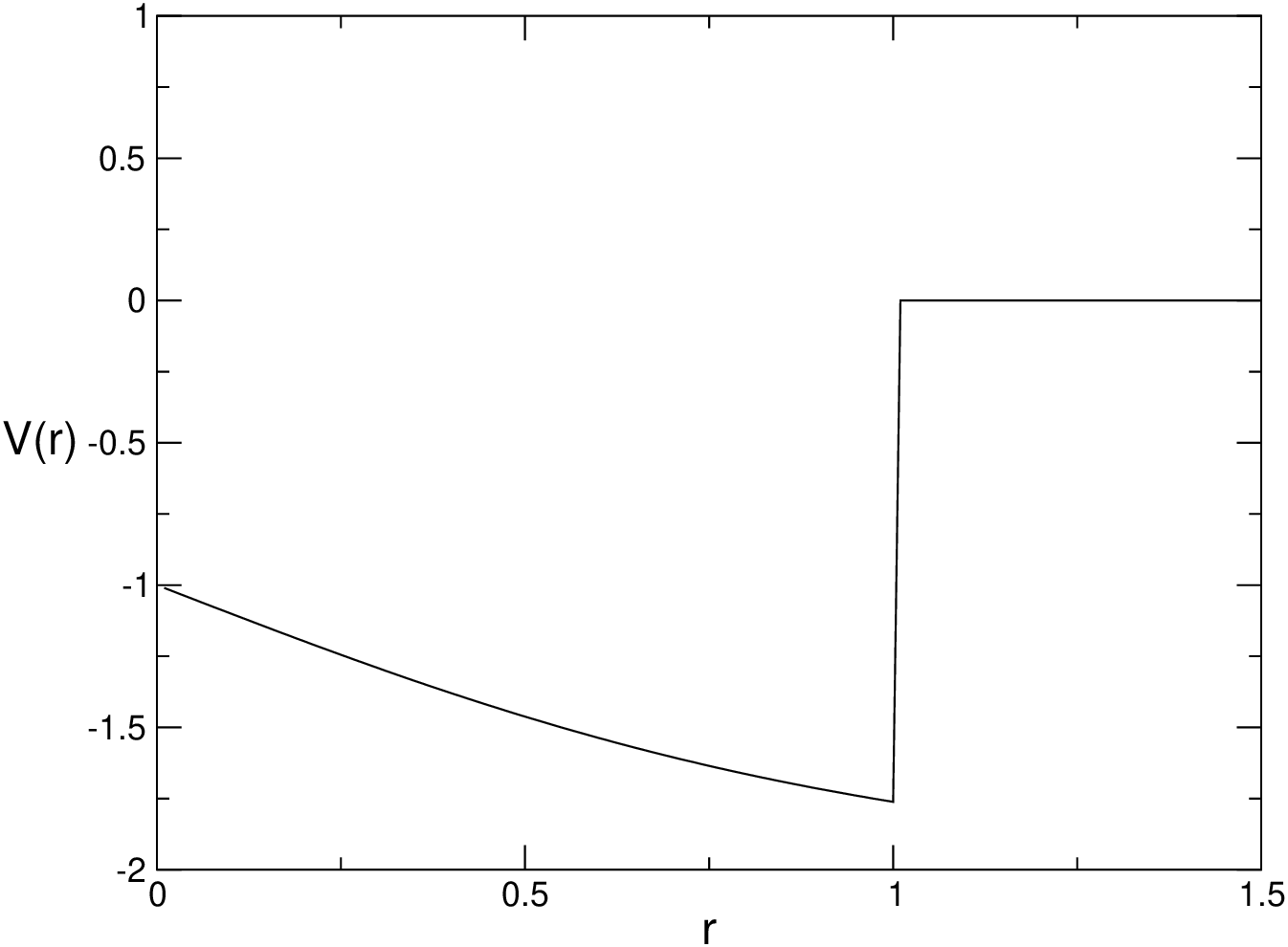}
\end{center}
\caption{Plot of the truncated hyperbolic tangent potential for
$D=-1$, $k=1$, $B=1$ and $a=1$ } \label{fig1}
\end{figure}

The boundary  conditions for our problem are
\begin{equation}
u_{l}(0)=0, \label{bound1}
\end{equation}
\begin{equation}
u_{l}(a)=u_{r}(a) \label{bound2}
\end{equation}
\begin{equation}
 u^{\prime}_{l}(a)=u^{\prime}_{r}(a) \label{bound3}
\end{equation}
where $u_{l}$ and $u_{r}$ are the solutions of Eq \eqref{kg} in
 $[0,a]$ and $[a,\infty)$, respectively. The prime indicates
a derivative with respect to the radial variable. The solutions
$u_{l}$ and $u_{r}$  are given
 by
\begin{equation}\label{sold}
\begin{split}
&u_{l}(r)=\frac{(\tanh(kr)-1)^{a}}{r}\Big(C_{1}F\left[c,d,f;\frac{1+\tanh
(kr)}{2}\right]\\
&+C_{2}\left(\frac{\tanh(kr)+1}{2}\right)^{-2b}F\left[g,h,j;\frac{1+\tanh
(kr)}{2}\right]\Big)(1+\tanh(kr))^{b}
\end{split}
\end{equation}
where
\begin{align*}
a&\equiv a\left(m,D,E,B,k\right)=\frac{\sqrt{m^{2}-(E+B)(E+B-2D)-D^{2}}}{2k}\\
b&\equiv b\left(m,D,E,B,k\right)=\frac{\sqrt{m^{2}-(E+B)(E+B+2D)-D^{2}}}{2k}\\
c&\equiv
c\left(m,D,E,B,k\right)=\frac{k-\sqrt{k^{2}-4D^{2}}}{2k}+b\left(m,D,E,B,k\right)+a\left(m,D,E,B,k\right)\\
d&\equiv
d\left(m,D,E,B,k\right)=\frac{k+\sqrt{k^{2}-4D^{2}}}{2k}+b\left(m,D,E,B,k\right)+a\left(m,D,E,B,k\right)\\
f&\equiv f\left(m,D,E,B,k\right)=1+2b\left(m,D,E,B,k\right)\\
g&\equiv
g\left(m,D,E,B,k\right)=\frac{k-\sqrt{k^{2}-4D^{2}}}{2k}-b\left(m,D,E,B,k\right)+a\left(m,D,E,B,k\right)\\
h&\equiv
h\left(m,D,E,B,k\right)=\frac{k+\sqrt{k^{2}-4D^{2}}}{2k}-b\left(m,D,E,B,k\right)+a\left(m,D,E,B,k\right)\\
j&\equiv j\left(m,D,E,B,k\right)=1-2b\left(m,D,E,B,k\right)
\end{align*}
and
\begin{equation}\label{solf}
u_{r}(r)=C_{3}\,\frac{e^{-k\,r}}{r}\, ,
\end{equation}
where $k^2=m^{2}-E^{2}$.

Making use of the boundary conditions (\ref{bound1}),
(\ref{bound2}), and (\ref{bound3}) we obtain an eigenvalue energy
equation  in terms of $B,D,k$ and $a$; that is, $E\equiv
E\left(B,D,k,a\right)$. It should be mentioned that, if we impose
the condition $D=0$  in the eigenvalue equation, we recover the
equation governing the energy levels for a square well.

\begin{figure}[tbp]
\begin{center}
\includegraphics[width=10cm]{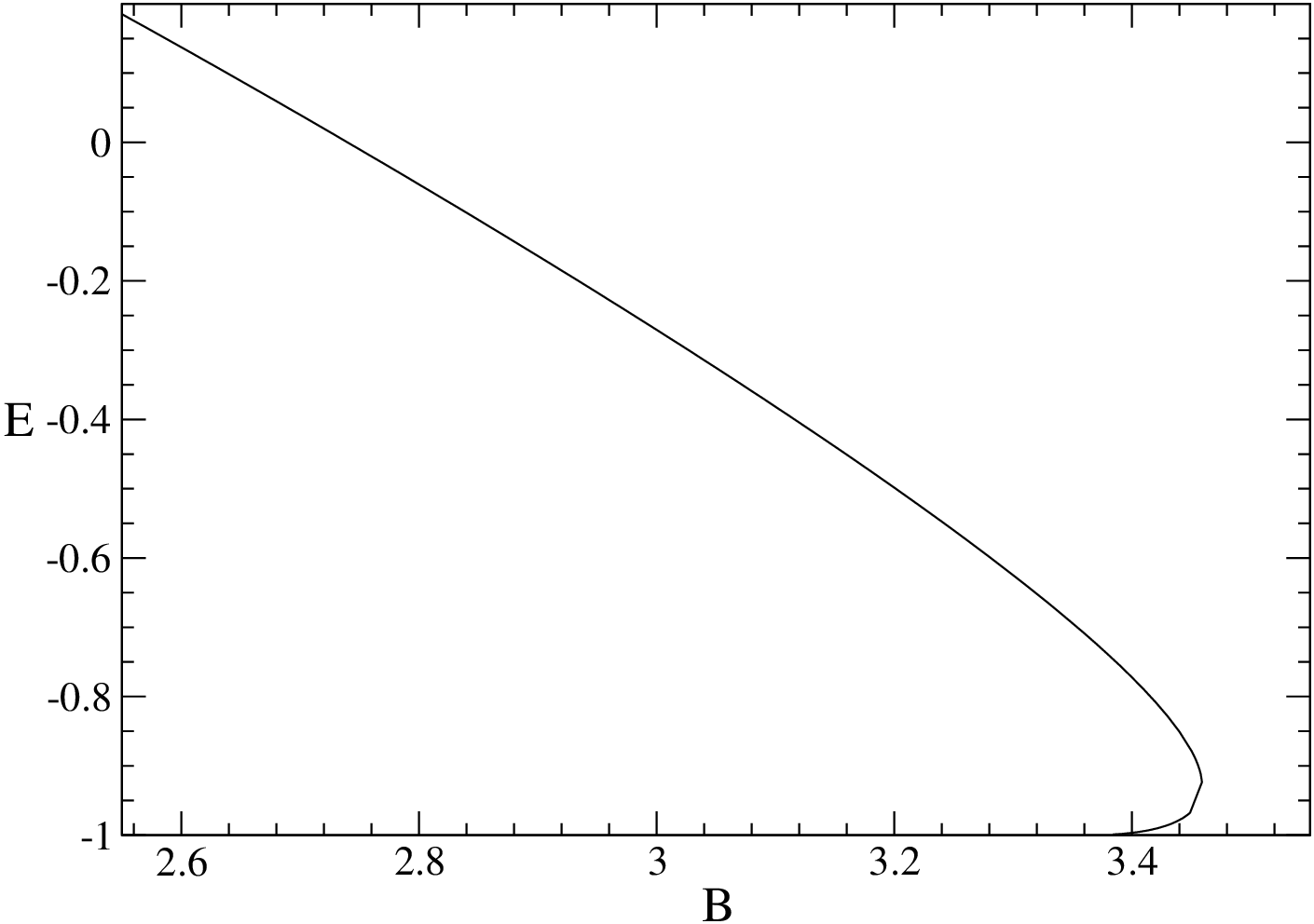}
\end{center}
\caption{Spectrum of the Klein-Gordon equation with $B$ between 2.54
and 3.45, $D=$ 0.86, $k=1$, $a=1$} \label{fig2}
\end{figure}

Now we proceed to illustrate the behavior of the eigenvalue energy
equation  for different values of the parameters  $B,D,k$ y $a$: In 
Fig. (\ref{fig2}) we observe the Schiff-Snyder-Weinberg effect, where 
antiparticles states emerge from the lower continuum. This phenomenon can 
be better understood noticing that the Klein-Gordon  norm \cite{Greiner,Fulling}

\begin{equation}
N=2\int dx^3*(E-eA_{0})\phi^2
\label{norma}
\end{equation}
is not positive definite. Particle and antiparticle states are identified, with 
the exception of the accidental case $N=0$,  according to the sign of the 
norm $N$ (\ref{norma}). 

\begin{figure}[tbp]
\vspace{0.2cm}
\begin{center}
\includegraphics[width=10cm]{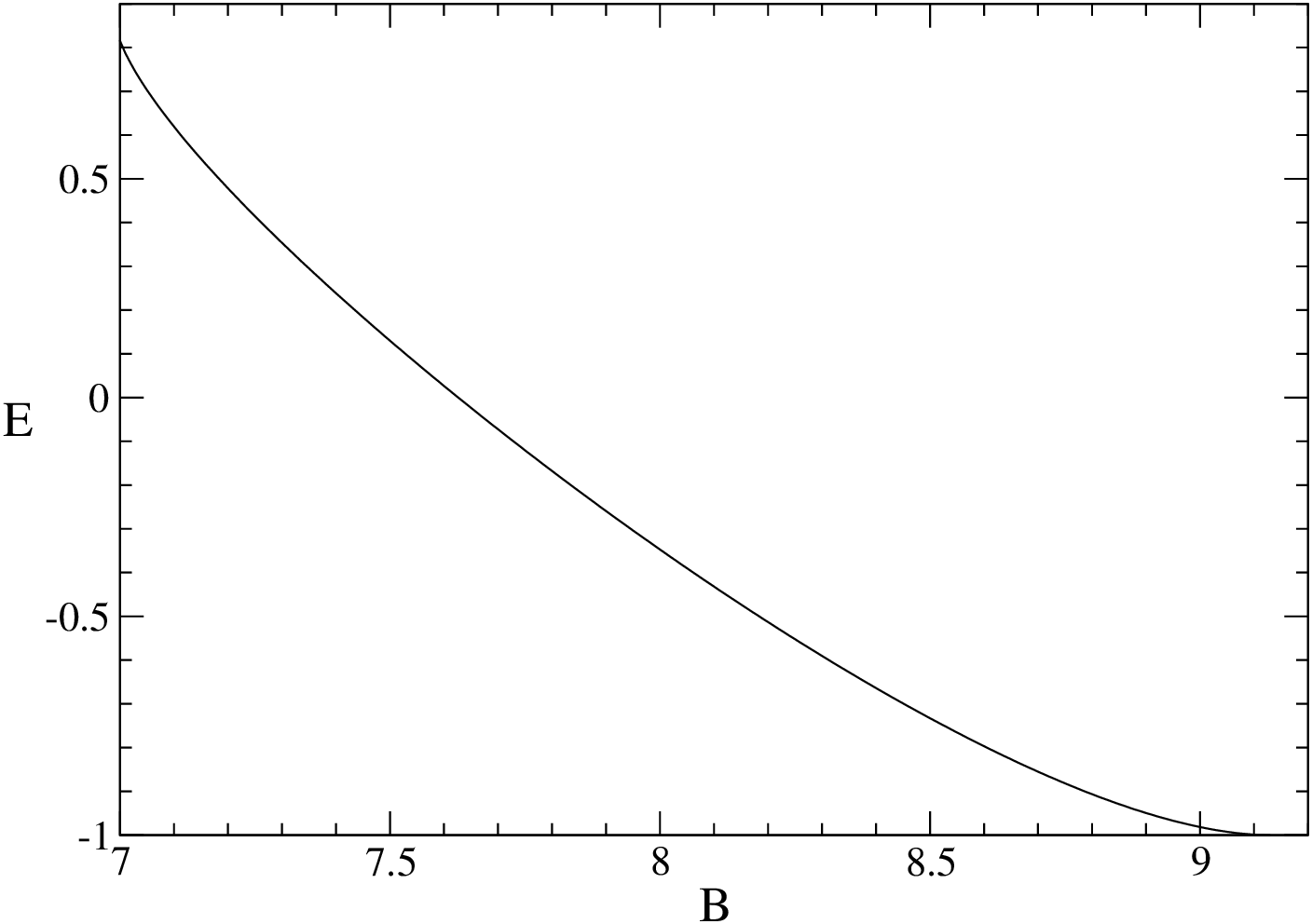}
\end{center}
\caption{Spectrum of the Klein-Gordon equation with $B$ between 6.99
and  9.13, $D=-11$, $k=10$, $a=0.6$} \label{fig3}
\end{figure}

In contrast to the Schiff-Snyder effect case, Fig (\ref{fig3}). shows an embedding
of the energy levels into the negative continuum.

We observe that in the situation described by Fig. (\ref{fig2}),  we
obtain complex energy values for values of  $B$ larger than the
turning point $B=3.459$ (for this critical value  the energy reaches
the value $E=-0.92355$ and starts generating antiparticle bound states
coming form the negative energy continuum). The associated
eigenstate with this critical value has a zero norm in the sense of 
Ref \cite{BL:79} and Eq. (\ref{norma}). Below this critical value, the norm
associated with the energy states remains negative and above this
value it becomes positive.

For the situation described in Fig. (\ref{fig3}) the energy $E$
reaches the value $E=-1$ for $B=9.1305$. In this case, the norm
remains always positive. Beyond this critical value, there is an
embedding process into the negative energy continuum.

Making $k^2=E^{2}-m^{2}\equiv-\kappa^{2}$ in Eq. \eqref{solf}, we
obtain  $u_{f}(r)$ for the  $s$ waves belonging to the continuum:
\begin{equation}\label{solfc}
u^{c}_{r}(r)=C_{3}\,\frac{e^{i\left(-\kappa\,r+\delta\right)}}{r}
\end{equation}
Repeating the procedure applied in the derivation of the eigenvalue
equation with this new function we obtain an expression giving
$\delta$ in terms of the energy, that is:
$\delta\left(E\left(B,D,k,a\right)\right)$.

We now show the figures depicting
$\delta\left(E\left(B,D,k,a\right)\right)$ for the cases described
by Fig. (\ref{fig2}) and Fig. (\ref{fig3}) respectively.

In order for a resonance to exist the Wigner time delay must be
positive. This is defined as

\begin{equation}
\tau=2\frac{d\delta}{dE},
\end{equation}
so the phase shift must increase through $\pi/2$ with the energy. 
A decrease in the
phase shift as the energy increases through $\pi/2$ produces an unphysical
negative time delay \cite{Kamke}, and does not constitute a
resonance.

\begin{figure}[tbp]
\vspace{0.5cm}
\begin{center}
\includegraphics[width=10cm]{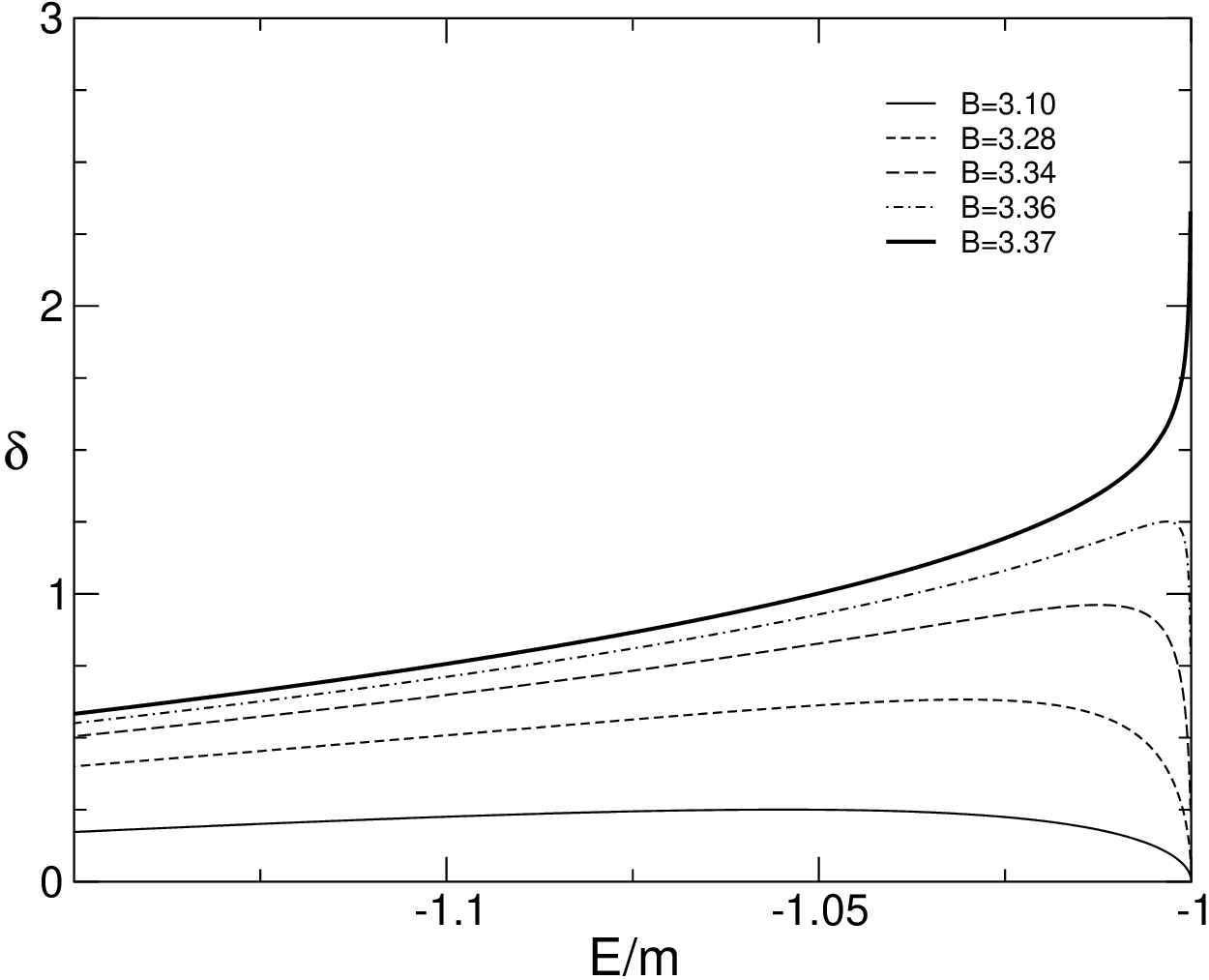}
\end{center}
 \caption{Phase shift associated with the energy
spectrum described in Fig. (\ref{fig2}).} \label{fig4}
\end{figure}

In Fig. (\ref{fig4}) the curves  tend to the asymptotic value
$\pi/2$ for $\delta$ without reaching it; here we are in the
presence of an attempt  of formation of a bound  state but the
potential is not strong enough in order to support such a state.
Nevertheless, we are not in the presence of a resonance since, for
this situation,  the Wigner time delay \cite{EW:55} is not positive.
Fig. (\ref{fig4}) shows that the appearance of a antiparticle state,
phenomenon that takes place for the value $B\approx3.459$, and that
does not represent a dramatic change in the phase shift (a jump to
$\delta=\pi$) that experiences a bound state crossing through
$E=-1$.

\begin{figure}[tbp]
\begin{center}
\includegraphics[width=10cm]{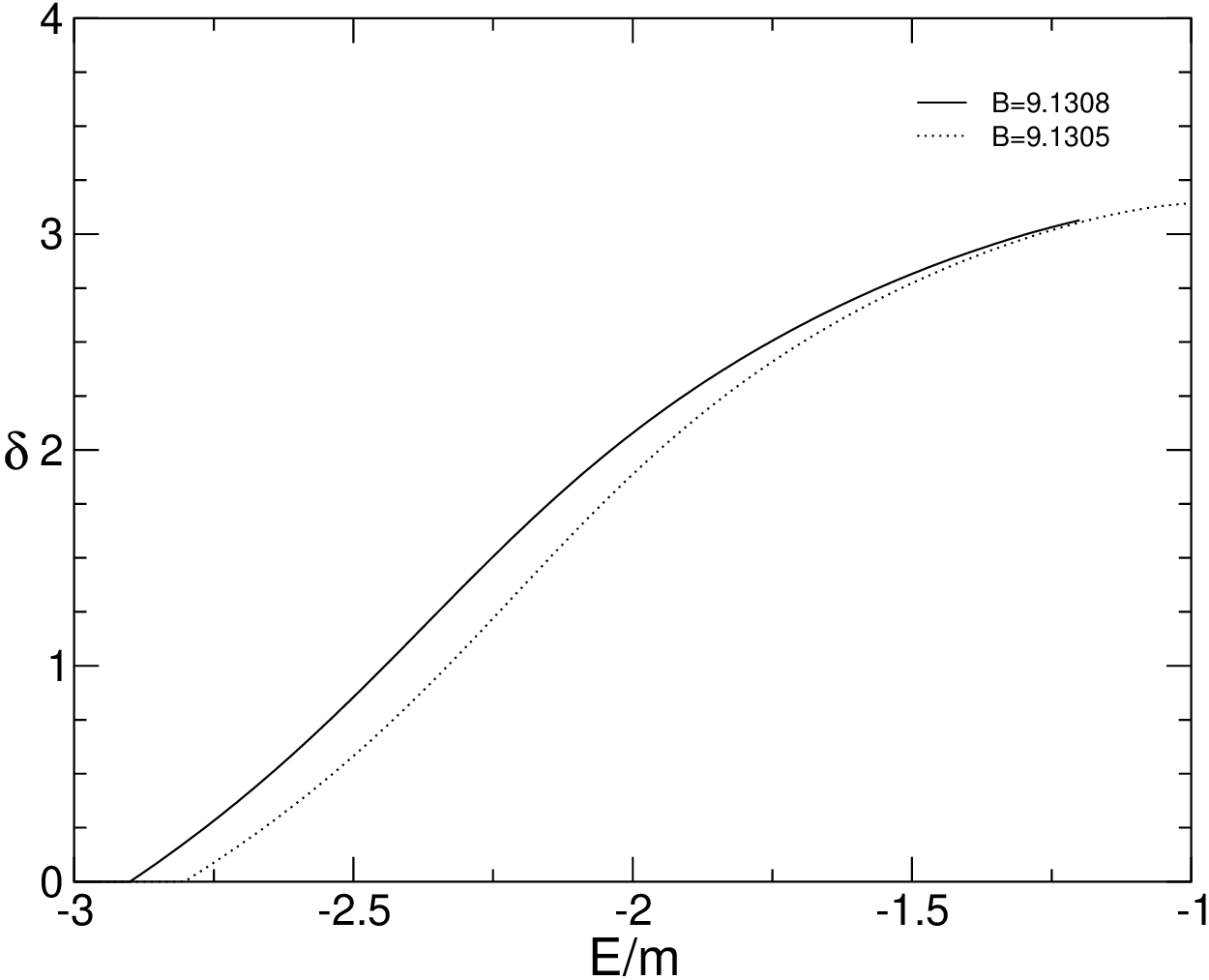}
\end{center}
\caption{Phase shift associated with the energy spectrum described
in Fig. (\ref{fig3}) } \label{fig5}
\end{figure}

In Fig. (\ref{fig5}) the curves overpass the value  $\pi/2$,
reaching the value of $\pi$ for  $\delta$, finding us in this way in
the presence of a resonance for the values of $B$ shown in the
figure. It is worth noticing that in this case the Wigner time delay
is positive. The phase shifts experiences a drastic change in the
resonant value.

From Fig (\ref{fig2}) and Fig. (\ref{fig3})  we observe that both
curves possess different concavity, being positive in the
resonant-embedding regime and negative when the Schiff-Snyder effect
is present. This behavior is analogous to the one observed for $p$
states in the scalar well \cite{BL:79}. It is worth mentioning that
both the Schiff-Snyder and the embedding effect  depend
on the sign of the shape constant $D$ in the potential $V(r)$
(\ref{V}). 

We have shown evidence of a short-range potential that for $s$ waves 
exhibit 
Fig. (\ref{fig3}) and Fig. (\ref{fig5}) show evidence that, for
$s$ waves, there is a resonant behavior for a class of short range
potentials, with $D<0$  in Eq. (\ref{V}), this result is, based on the 
results reported in Ref.\cite{VP:71}, new and unexpected.

\acknowledgments{We thank Dr. Ernesto Medina for reading and improving the manuscript. 
This work was supported by FONACIT under project G-2001000712. }

\end{document}